# Not-so-simple stellar populations in nearby, resolved massive star clusters

Richard de Grijs[1,2] and Chengyuan Li[3]

[1]*Kavli Institute for Astronomy & Astrophysics and Department of Astronomy, Peking University, Yi He Yuan Lu 5, Hai Dian District, Beijing 100871, China*
[2]*International Space Science Institute–Beijing, 1 Nanertiao, Zhongguancun, Hai Dian District, Beijing 100190, China*
[3]*Department of Physics and Astronomy, Macquarie University, Balaclava Road, North Ryde, NSW 2109, Australia*

**Abstract** – Around the turn of the last century, star clusters of all kinds were considered 'simple' stellar populations. Over the past decade, this situation has changed dramatically. At the same time, star clusters are among the brightest stellar population components and, as such, they are visible out to much greater distances than individual stars, even the brightest, so that understanding the intricacies of star cluster composition and their evolution is imperative for understanding stellar populations and the evolution of galaxies as a whole. In this review of where the field has moved to in recent years, we place particular emphasis on the properties and importance of binary systems, the effects of rapid stellar rotation, and the presence of multiple populations in Magellanic Cloud star clusters across the full age range. Our most recent results imply a reverse paradigm shift, back to the old simple stellar population picture for at least some intermediate-age (~1–3 Gyr-old) star clusters, opening up exciting avenues for future research efforts.

## 1. Simple or not?

Since the launch of the *Hubble Space Telescope* (*HST*), more than 20 years ago now, studies of 'populous' star clusters (that is, clusters with masses in excess of $\sim 10^4$ $M_\odot$ and of any age) have unveiled major new insights. The concept of star clusters as 'simple' stellar populations, that is, composed of stars of approximately the same age and the same chemical abundance (since all stars were thought to have formed roughly at the same time from the same progenitor molecular cloud), is now no longer viable. Our understanding of star cluster formation and evolution has been revolutionized (and has also become confused) by (i) old Milky Way 'globular' clusters that exhibit anticorrelations between the sodium and oxygen abundances of their member stars or star-by-star chemical variations (Kraft 1994; Gratton et al. 2001, 2004; Carretta et al. 2009); (ii) clusters of any age displaying multiple sequences in the diagnostic Hertzsprung–Russell diagram (relating stellar luminosities to their surface temperatures; e.g., Piotto et al. 2007; Villanova et al. 2007; Mackey et al. 2008; Milone et al. 2008, 2017), at any age; and (iii) the significant widths of the so-called main-sequence turn-off regions in many star clusters as young as a few tens of millions of years (e.g., Milone et al. 2009).

Simple stellar populations were always predominantly a theoretical construct, which—at best—could be closely approximated but never fully matched by real observations. Colour–magnitude diagrams (the observational counterparts of the theoretical Hertzsprung–Russell diagrams) showing sharp, well-defined main-sequence turn-offs imply that cluster member stars closely resemble single-age populations; tight main-sequence ridge lines suggest that any spread in their chemical composition is minimal, so that the only free parameter determining the morphology from the bottom of the main sequence to the turn-off region is stellar mass.

Yet, even those globular clusters that most closely match the simple stellar population paradigm exhibit deviations from this theoretical construct, which are unavoidable given that the theoretical concept applies to single stars only; realistic stellar populations tend to contain numerous binary systems. Imaging data are limited by the resolution of one's telescope, so that tight binary systems—that is, binaries with close separations of their component stars—might end up in a single resolution element. As a consequence, binary systems will appear brighter than single stars, simply by virtue of the additional stellar luminosity from the secondary component. At maximum, the effect of observing a binary system as a single luminosity component is a magnitude enhancement by ~0.75 magnitudes, or a factor of two in luminosity for equal-mass, and hence equal-luminosity, main-sequence–main-sequence binaries. The resulting observational colour–magnitude diagram will include a scattering of stars above (that is, brighter than) the single-star main sequence, filling in the parameter space between the single-star and equal-mass binary main sequences (see, e.g., Elson et al. 1998).

A second common deviation from the simple stellar population model, which is seen predominantly (but not exclusively) in old globular clusters, takes the form of a scattering of stars above (i.e., at brighter luminosities) the main-sequence turn-off, filling part of the parameter space between the main-sequence turn-off, the subgiant branch, and the horizontal branch. The age of a cluster's bulk stellar population is determined by the luminosity of its main-sequence turn-off, which is in essence a stellar-mass-dependent parameter: more massive stars exhaust hydrogen-to-helium nuclear fusion in their cores more rapidly than their lower-mass counterparts, so that they will evolve off the main sequence and become red-giant stars earlier. This implies that for a single-aged cluster population, there should be no main-sequence stars in a cluster's colour–magnitude diagram at brighter magnitudes than the main-sequence turn-off. Nevertheless, many populous clusters contain small populations of such stars. These objects, known as 'blue straggler stars,' are also products of binary systems; they are thought to have formed through either the transfer of mass from one component to the other, or by direct stellar collisions and subsequent mergers. Both processes cause stirring up of the stellar atmospheres of the resulting objects, thus making them appear 'rejuvenated,' suggesting younger ages.

## 2. Evolutionary aspects

To make matters more complicated, the effects of binary systems, blue stragglers, multiple sequences in the colour–magnitude diagrams, and a variety of other features associated with the 'not-so-simple' aspects of stellar populations are to some extent a function of the population's age. And this is where observations with the *HST* become crucial. Our Milky Way contains a population of around 160 known massive star clusters, with masses between about $10^4$ M$_\odot$ and a few $\times 10^6$ M$_\odot$, but these 'globular' clusters are exclusively old, with ages mostly greater than 10 Gyr. A number of younger, reasonably massive clusters are known to exist in the Milky Way, including NGC 3603, Westerlund 1, and the so-called red-supergiant clusters near the Galactic Centre (e.g., Davies et al. 2007, 2012), but they are often difficult to observe with current facilities owing to their locations in the Milky Way, behind significant layers of foreground dust. In addition, their masses of ~$10^4$ M$_\odot$ at ages of a few million years imply that these objects are unlikely to evolve into clusters similar to the current population of Galactic globular clusters by the time they reach similar ages: mass loss caused by stellar winds associated with advanced stellar evolutionary stages, stripping of their outermost stars by the prevailing Galactic tidal field, bulge or disk shocking, or internal two-body relaxation processes and the associated ejection of stars owing to interactions with binary systems in their cores will together cause these clusters to lose a significant fraction of their mass over the next 10 Gyr of evolution.

To explore any age dependence, and hence evolutionary aspects, we will thus need to resort to observations of young and intermediate-age (~1–3 Gyr-old) populous star clusters elsewhere, i.e., beyond the Milky Way. The most suitable environments to look for such clusters while retaining sufficient spatial resolution to still be able to distinguish the individual stars are represented by the Large and Small Magellanic Clouds (LMC and SMC). Numerous observations are readily available in the *HST* data archive of LMC and SMC clusters spanning the entire age range from the very young R136 cluster in the 30 Doradus star-forming region to the oldest globular-like clusters.

In Hu et al. (2010), we used such a data set to determine the global fraction of binary systems in the 15–25 Myr-old populous LMC cluster NGC 1818. At the distance of the LMC, for a distance modulus of $(m–M)_0 = 18.49$ mag or approximately 50 kpc (Pietrzyński et al. 2013; de Grijs et al. 2014), we cannot resolve binary systems into their individual components. As such, in colour–magnitude diagrams they will occupy the parameter space between the single-star and equal-mass binary sequences, with the actual locus of a given binary depending on the ratio of the component masses.

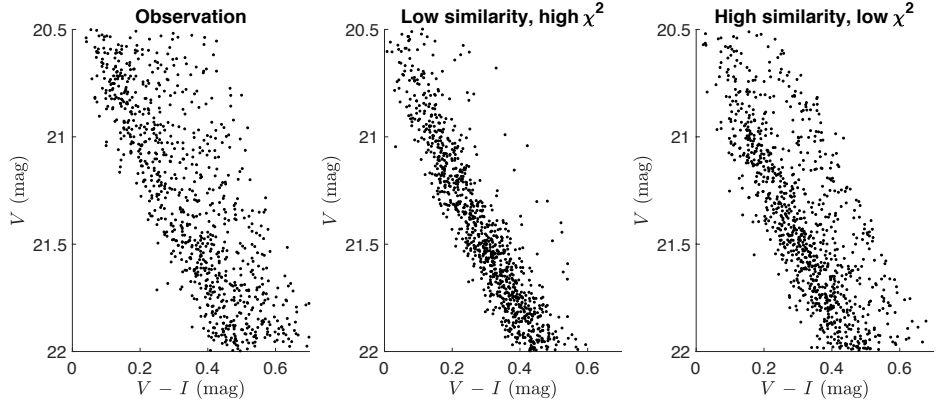

**Figure 1**: Illustrative example of our minimum-$\chi^2$ modelling. (*left*) Section of the main sequence of the young LMC cluster NGC 1805; (*middle, right*) Model colour–magnitude diagrams for main-sequence–main-sequence binary fractions of 5% and 50%, respectively.

We focused specifically on the shallowest part of the cluster's main sequence, where the effects of the presence of unresolved binary systems could be determined most easily, corresponding to the stellar mass range from 1.3 $M_\odot$ to 1.6 $M_\odot$ (F-type stars). The most robust approach to determining binary fractions based on colour–magnitude diagrams is by generating artificial colour–magnitude diagrams of clusters characterized by the same age and chemical composition ('metallicity') as one's target cluster, but with different fractions of binary systems. The next steps then consist of comparing the observational data with the model diagrams by application of $\chi^2$ minimisation. In essence, this process consists of quantification of the residuals of the comparison between the observational data and a given model representation by determining the $\chi^2$ value, as shown in Fig. 1. The best-fitting model is found when the $\chi^2$ value reaches a minimum: see Fig. 2. To properly model the characteristics of the binary population, one needs to adopt a prescription for the mass-ratio distribution, $q = m_s/m_p \leq 1$ (where $m_p$ and $m_s$ are the masses of the primary and secondary stellar component, respectively), i.e., $dN/dq \propto q^{-\alpha}$. Here, $\alpha$ usually varies between –0.5 and +0.5, but since observationally we can only distinguish binary systems with $q \geq 0.55$ from single stars, the precise value adopted for $\alpha$ does not make a significant difference in practice (Li et al. 2013)

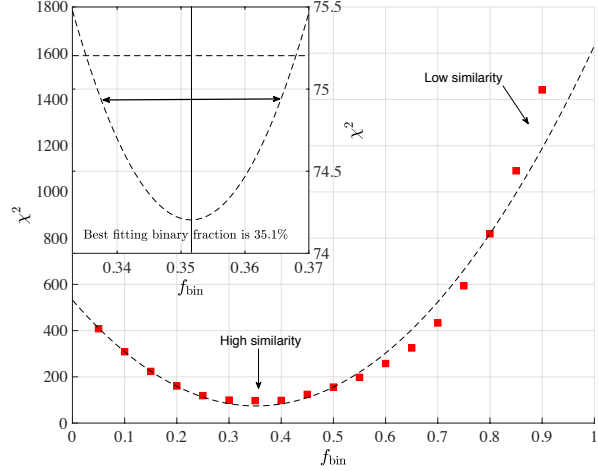

**Figure 2**: Determination of the best-fitting binary fraction based on $\chi^2$ minimisation ($y$ axis) of models characterized by a range in binary fractions ($x$ axis). (*inset*) Illustration of the 1σ uncertainty boundaries.

On the basis of numerous Monte Carlo experiments, in Hu et al. (2010) we eventually derived a best-fitting global binary fraction of 0.55 to unity, depending on the mass-ratio distribution at low $q$. For comparison, the binary fractions routinely obtained for old Galactic globular clusters are $\leq 10\%$. The difference with the binary fraction we determined for our young LMC cluster can be understood in the context of the evolution of binary systems in star clusters. Early-type field stars in the solar neighbourhood are characterized by binary fractions as high as 80% (Kouwenhoven et al. 2005), although F-type primary stars may have fewer companions. Nevertheless, F-type field stars still exhibit significantly higher binarity levels than old globular clusters; after all, it is fairly well-established that most stars with masses greater than 0.5 $M_\odot$ form in binaries or higher-order multiples (Kouwenhoven et al. 2005, 2007; Raghavan et al. 2010). Binary systems in globular clusters have undergone >10 Gyr of dynamical processing of their primordial binary systems; interactions between binaries and between binaries and single stars lead to either disruption, hardening, or exchange encounters. Generally speaking, Heggie's law applies: 'hard binaries tend to get harder, while soft binary systems become softer or are disrupted' (Heggie 1975). It is, therefore, not surprising that globular cluster binarity levels are very low.

Reflecting on earlier indications that NGC 1818, as well as other young populous clusters in the LMC, showed clear evidence of mass segregation (e.g., de Grijs et al. 2002a,b), we proceeded to explore whether this was mirrored by the distribution of the binary systems in the cluster (de Grijs et al. 2013; Li et al. 2013). One might perhaps expect dynamical processes to act similarly upon tightly bound binary systems as they would on single stars with total masses equivalent to the binary systems' combined masses. Since dynamical processing of stars spanning a mass range in a given star cluster leads to mass segregation, similar reasoning would suggest that the binary systems would eventually also become more centrally concentrated than the single stars with masses similar to those of the binaries' primary components. This is not what we found in NGC 1818. In fact, we found the opposite behaviour: the cluster exhibited a dearth of binary systems within its core compared with its periphery.

Initially, we speculated that we might have uncovered evidence of the disruption of 'soft' binary systems—which are less tightly gravitationally bound than their 'harder' counterparts characterized by smaller separations—on timescales of a few tens of millions of years. We proceeded to verify this suggestion by taking a two-pronged approach. First, we applied identical analysis techniques to the similarly young but much more compact LMC cluster NGC 1805. Indeed, this object behaved as expected: it exhibited a higher fraction of main-sequence–main-sequence binary systems in its core compared with that in the cluster's outer regions (Li et al. 2013). NGC 1805 and NGC 1818 were selected to be as similar to one another as possible in terms of their physical characteristics (masses, ages, metallicities, tidal effects as determined by their distances from the LMC's centre, etc.), with as only key difference their (linear) sizes, for which we took their core or half-light radii as proxies. Given that the much more compact cluster NGC 1805 indeed showed a higher fraction of binaries in its core than at larger radii, this provided support to our suggestion that while both clusters had similar chronological ages (stellar ages), they had very different *dynamical* ages as expressed in units of their core or half-mass (half-light) relaxation timescales.

Second, we proceeded to model both clusters using *N*-body (numerical) simulations that reflected the clusters' reality as closely as possible (Geller et al. 2013, 2015). We started our simulations of our ~$2\times10^4$ $M_\odot$ clusters ($N$ = 36,000) from both smooth (Plummer sphere) and clumpy (highly fractal) stellar distributions, without gas, and evolved both configurations to an age of 30 Myr, starting from a constant binary fraction at any cluster radius. The clumpy distribution transitions into a smooth, Plummer-sphere-like distribution on timescales of order 30 Myr (Geller et al. 2013), the approximate age of our young clusters. And indeed, the difference in initial radii and the associated difference in the clusters' crossing times was found to be responsible for the observed difference in the radial behaviour of the clusters' binary fractions. Interestingly, after an initial period during which the soft binary systems are preferentially disrupted, dynamical mass segregation takes over and the resulting 'minimum' in the fraction moves radially outwards over time—thus suggesting that we may potentially use the radial distribution of a cluster's binary fraction as a dynamical clock.

## 3. Complications at ages in excess of a few hundred thousand years

Our next step in the evolutionary sequence focuses on a pair of LMC clusters with ages of a few $\times10^8$ yr, NGC 1831 and NGC 1868. Both clusters exhibit clearly broadened main sequences and main-sequence turn-off regions (e.g., Li et al. 2014a). Although part of this broadening can be explained by the presence of a population of binary systems, that is only part of the story. In recent years, a number of different explanations have been proposed to account for broadened features in cluster colour–magnitude diagrams, including (i) a spread in helium abundances, (ii) a range in ages and/or metallicities of a cluster's member stars, (iii) the presence of a population of rapidly rotating stars, and (iv) stellar variability. Of these options, we can easily rule out the presence of a population of helium-enhanced stars, because such stars would occupy secondary main sequences that are *bluer* than that of the bulk of the stellar population; the broadened main sequences tend to be broadened towards redder colours instead. We will return below to the recent suggestion that stellar variability, particularly of so-called δ Scuti stars—low-mass variable stars located in the colour–magnitude diagram where the main sequence intersects with the 'instability strip'—could be responsible for at least some of the broadening.

This thus leaves us with three possibilities for the observed broadening of the stellar evolutionary features. Fortunately, the contributions of these effects can be disentangled through careful analysis of cluster colour–magnitude diagrams. First, we can use the lower main sequence, well away from the main-sequence turn-off, to determine a cluster's global binary fraction in a similar way as we did for NGC 1805 and NGC 1818 above. For both NGC 1831 and NGC 1868, this results in binary fractions of order 30% for $q > 0.6$. Extrapolating to the lowest-mass-ratio regime, this implies global binary fractions of 63% and 76% for NGC 1831 and NGC 1868, respectively (Li et al. 2014a).

Second, while the broadened main-sequence turn-off regions in both clusters can be bracketed by stellar isochrones of different ages, spanning an age range of ~300 Myr in both cases, we can use the vertical extent (in magnitude) of the clusters' 'red clumps' to constrain the maximum possible age (or metallicity) range among their member stars. In Li et al. (2014a), we found that the maximum possible age ranges allowed by the extents of the main-sequence turn-offs are 380 Myr and 350 Myr for NGC 1831 and NGC 1868, respectively. The extents of the red clumps tell a different story, however, in the sense that such a large age range could be encompassed by this feature in NGC 1868, but it is ruled out for NGC 1831.

This thus requires an alternative explanation for at least part of the extended main-sequence turn-off in NGC 1831, beyond the effects of a population of binary systems and possibly a small age range. Such an explanation is found in the properties of rapidly rotating stars, with masses $> 1.2$ $M_\odot$ (e.g., Georgy et al. 2014), within the cluster's bulk population. Rapidly rotating stars can retain hydrostatic equilibrium at lower temperatures than their more slowly or non-rotating counterparts, thus leading to redder colours, and thus producing a broadening of the main-sequence turn-off of intermediate-age (~1–3 Gyr-old) clusters. We did not discuss the effects of rapidly rotating stars for our two young clusters, because the effects in the resulting colour–magnitude diagram of the reduced temperature are small or negligible for the hotter, younger (A- and early B-type) stars dominating the stellar populations in these clusters: see Fig. 3.

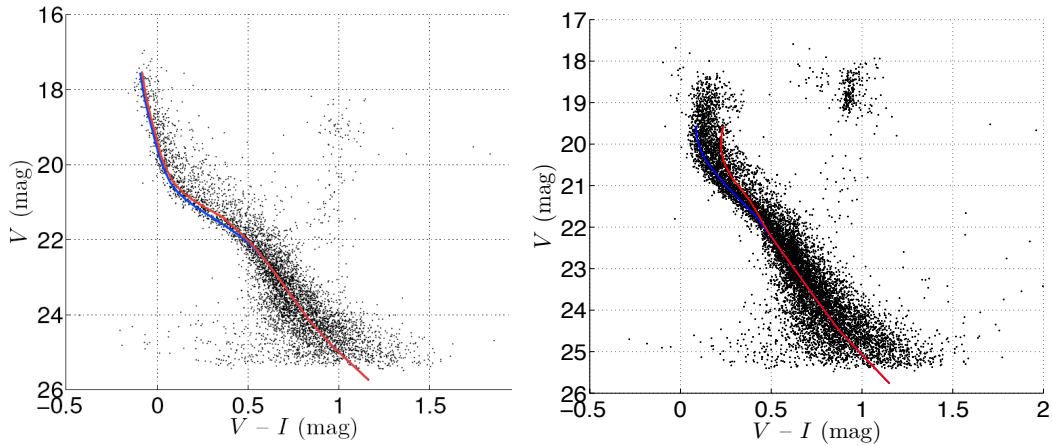

**Figure 3**: Colour–magnitude diagrams of (*left*) NGC 1818 (a few $\times 10^7$ yr old) and (*right*) NGC 1831 (a few $\times 10^8$ yr old). Blue: no stellar rotation; red: rotation at 55% of the critical break-up rate. Both colour–magnitude diagrams represent the respective cluster populations after corrections for field-star contamination based on statistical background subtraction (for an example of this process, see Section 5 below).

To verify the suggestion that a range in stellar rotation rates may well be at the root of the extended main-sequence turn-offs observed for NGC 1831 and NGC 1868, we proceeded to model their effects on the basis of extensive Monte Carlo simulations (Li et al. 2014a). However, in order to do so, we need to know how stellar rotation rates are distributed in realistic star clusters with ages and metallicities similar to those of our sample objects. Such information, however, is at present unavailable. The most appropriate current data set that we could employ as the basis for our modeling is provided by Royer et al. (2007), whose distribution of the ($v \sin i$) values (where $v$ is the rotational velocity and $i$ represents the inclination) of B9 to F2-type stars shows characteristic peaks at rotation rates of approximately 0.10 and 0.55 times the rotational break-up rate, corresponding to non- or slowly rotating stars and their complement of rapidly rotating counterparts, respectively.

For each cluster, we first used a Monte Carlo approach to generate a simple stellar population characteristic of the cluster's bulk properties. Next, we randomly assigned stellar rotation rates to cluster stars with masses > 1.2 $M_\odot$, for which the effects of rotation become apparent in colour–magnitude diagrams. We then assigned 'binary' status to 70% of the artificial stars (corresponding to the approximate binary fractions pertaining to our sample clusters) and adjusted their luminosities to reflect their binary nature (adopting a flat mass-ratio distribution); this represents the approximate global binary fractions of both sample clusters. Finally, we added representative uncertainties (obtained from the actual observational data for the relevant magnitude range) to the artificial stars' colours and magnitudes. Comparison of our artificial colour–magnitude diagrams with those observed, again in a minimum-$\chi^2$ sense (as quantified by consideration of the residuals of the fits), showed that implementation of a realistic stellar rotation distribution can indeed explain the observational data.

Let us now briefly return to stellar variability as a potential cause of the observed broadened main sequences. Salinas et al. (2016) recently suggested that a fraction of the main-sequence stars in the magnitude range where the main sequence intersects with the so-called instability strip (where pulsating δ Scuti variables are found) could contribute to a broadening of the main sequence if their magnitudes represent snapshots rather than average values over a full pulsation period. Thus far, the observational data used for every single cluster for which evidence of broadened main sequences has been published is, in essence, single-epoch data rather than period-averaged photometry. The importance of this effect depends on the 'incidence' of stellar variability, that is, the fraction of main-sequence stars in the instability strip that actually pulsate; not every single star will become a pulsating variable, with incidences ranging up to 50%. Interestingly, the ages where this effect is expected to be most notable range from approximately 1 to 2.5 Gyr—fully overlapping with the age range where we find extended main-sequence turn-offs in Magellanic Cloud star clusters (for a discussion, see also de Grijs 2017).

**4. Beyond the main sequence**

Clearly, the morphology of the main sequence and its turn-off alone is insufficient to conclusively identify the main cause of these broadened features in intermediate-age star clusters. However, additional inclusion in one's analysis of the next stellar evolutionary phase, that of the subgiant branch, has the potential to resolve the issues. In Li et al. (2014b), we showed for the first time that while a cluster's main-sequence turn-off region may be broad, the associated subgiant branch of NGC 1651 is so tight that it essentially constrains the cluster's stellar population to a very narrow age range; the concept of the 'simple' stellar population thus made a comeback: see Fig. 4.

We found that while the width of the main-sequence turn-off in the 2 Gyr-old massive ($\sim 1.7 \times 10^5$ $M_\odot$) cluster NGC 1651, both in the cluster core and for its full stellar population, could potentially be explained by adopting a range of stellar ages spanning 450 Myr, its subgiant branch constrained any likely age range to a maximum of ~80 Myr. Our main conclusion, that is, that the presence of an extended main-sequence turn-off does not necessarily imply a range in stellar ages, was shortly afterwards confirmed by Bastian & Niederhofer (2015). We tentatively suggested that this might be explained by the presence of a population of rapidly rotating stars, although we clarified that the final word had not yet been uttered about that latter issue.

Nevertheless, competing scenarios were proposed almost immediately, with suggestions that in our analysis we had neglected a scattering of stars above the subgiant branch (Goudfrooij et al. 2015). In fact, we had attributed those stars to binary systems of which one of the components was going through the subgiant phase, but Goudfrooij et al. (2015) argued that instead of rotation, the effects of atmospheric overshooting might be at the basis of our observations. In fact, in order to reconcile their models with the observational data, Goudfrooij et al. (2015) adopted a level of convective core overshooting, $\Lambda_c$, of up to $\Lambda_c = 0.5$ for subgiant stars; however, *Kepler* observations of ~1.3 $M_\odot$ subgiant stars are consistent with $\Lambda_c^{max} = 0.2$ (Deheuvels et al. 2016; see also Claret & Torres 2016 for results based on double-lined eclipsing binary systems).

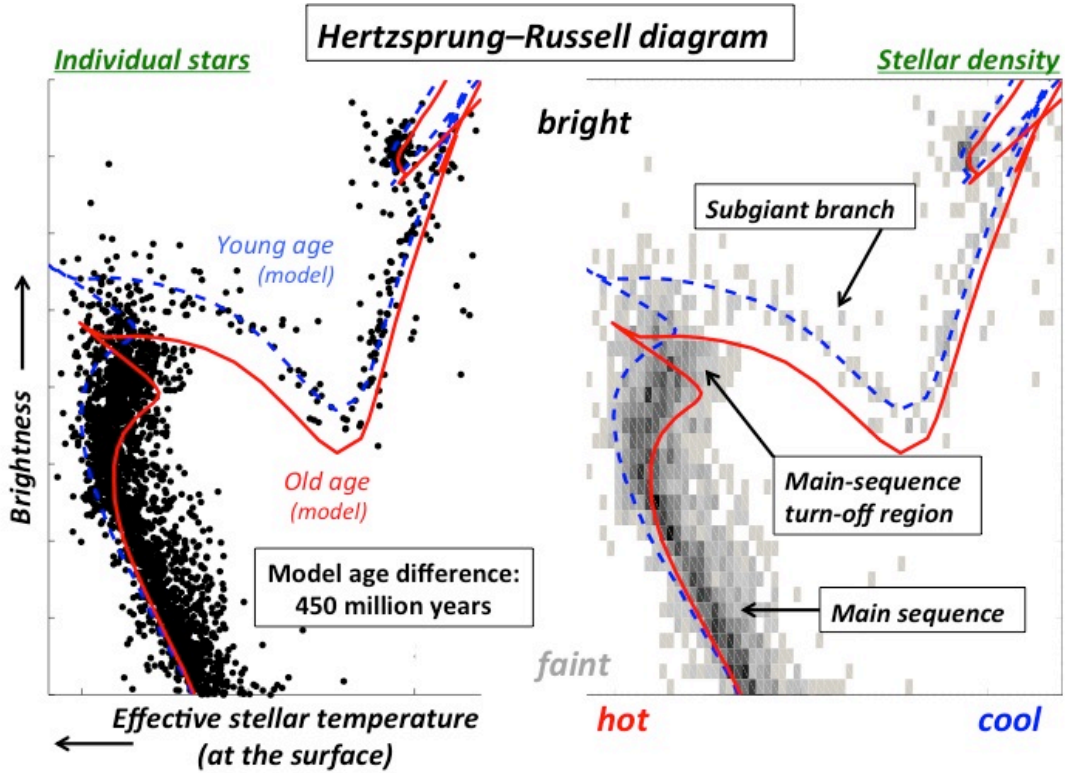

**Figure 4**: Colour–magnitude diagrams of NGC 1651. (*left*) Stars; (*right*) Hess diagram.

In addition, while we do not contest the suggestion that different amounts of overshooting would lead to changes in stellar luminosities, this process is an intrinsic physical process and we do not understand from basic physical principles why different stars in the same stellar population would exhibit different levels of overshooting; the process should act similarly for the entire population. Another, somewhat more troubling issue that was raised is related to the effects of increased stellar rotation on the lifetime of a main-sequence star. Naively speaking, at lower temperatures, equal-mass stars will have longer main-sequence lifetimes, which would directly affect their luminosities and, hence, produce a measurable width of the subgiant branch—but this is not seen in our colour–magnitude diagrams.

To put to rest any remaining suggestions that an age range could still be accommodated in some of the intermediate-age clusters exhibiting extended main-sequence turn-offs, we extended our exploration to the 1.6 Gyr-old SMC cluster NGC 411 (Li et al. 2016b). This object is one of the lowest-mass ($\sim 3.2 \times 10^4$ M$_\odot$) clusters exhibiting an extended main-sequence turn-off with a width that is among the broadest widths known; if the width were entirely owing to an age spread, this would imply an age range of between 700 Myr and 1 Gyr (Girardi et al. 2013; Goudfrooij et al. 2014). However, since clusters of these ages do not contain any sizeable gas reservoirs anymore (Bastian & Strader 2014), it is entirely unclear how star formation could have proceeded for such an extreme length of time. After a star cluster has been formed from its progenitor giant molecular cloud, it is cleared of any remaining gas within ~40 Myr owing to the combined effects of supernova- and stellar-wind-driven gas expulsion, thus leaving a purely stellar system. Although a variety of authors have suggested that new stars may form where the massive winds from evolved, asymptotic giant-branch stars intersect and collide, for this process to be viable the cluster's gravitational potential needs to be sufficient to retain that gas. Owing to the low mass of NGC 411, and hence the low escape velocity, it is unlikely that these boundary conditions are met in the cluster. Indeed, while its main-sequence turn-off is exceedingly broad, the cluster's subgiant branch clearly favours a single-aged stellar population, ruling out an extended star-formation history (Li et al. 2016b).

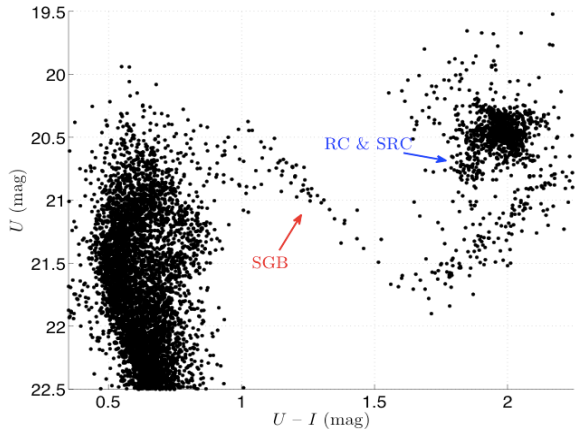

**Figure 5**: Colour–magnitude diagram of NGC 419, based on the $U$ band, where multiple populations tend to be more clearly visible than at redder wavelengths. SGB: subgiant branch; RC: red clump; SRC: secondary red clump. (Data courtesy of Antonino Milone)

Perhaps even stronger evidence of the stellar rotation scenario was provided by Wu et al. (2016) on the basis of their study of the 2.0 Gyr-old SMC cluster NGC 419. Exploring the maximum age ranges among the cluster's stars based on both the width of the main-sequence turn-off and that of its subgiant branch, we noticed that the red half of the cluster's subgiant branch represented a reasonably close match to the main-sequence turn-off, but that the subgiant branch's blue section was significantly narrower: see Fig. 5, where we use bluer filters than in Wu et al. (2016), which shows the effect even more clearly.

Intrigued, we explored the evolution of stellar rotation rates across the subgiant branch and concluded that, for the first time, we had uncovered evidence of rotational deceleration of stars evolving from the main-sequence turn-off to the red-giant branch. This had been expected on the basis of the physical law requiring conservation of angular momentum, but it had never been seen before. Our result thus conclusively showed that the subgiant branch width could be attributed to stellar rotation properties and that assuming an age range did not work.

**5. Multiple populations after all!**

Quite unexpectedly, while exploring the colour–magnitude features of the 1.7 Gyr-old LMC cluster NGC 1783, we noticed two additional, tight sequences of stars extending to well above the cluster's bulk main-sequence turn-off. Careful analysis implied that we had found evidence for the existence of two younger populations, with ages of 400 and 800 Myr, in a more evolved cluster (Li et al. 2016a). An extensive search of the *HST* data archive uncovered another two intermediate-age clusters which also exhibited younger populations alongside their bulk stellar populations, NGC 1806 in the LMC and NGC 411.

Although we suggested that these younger populations may have been formed from gas that had been swept up from their host galaxies' interstellar medium, we now believe that that interpretation is likely not correct. Instead, in Hong et al. (2017) we show, based on *N*-body simulations, that the younger populations in NGC 411 and NGC 1806 can be explained naturally from mergers of these massive clusters with younger, less massive clusters. This idea is not as far-fetched as it might seem, given that the Magellanic Clouds are known to host numerous binary star clusters (Bica et al. 1999; Dieball et al. 2002).

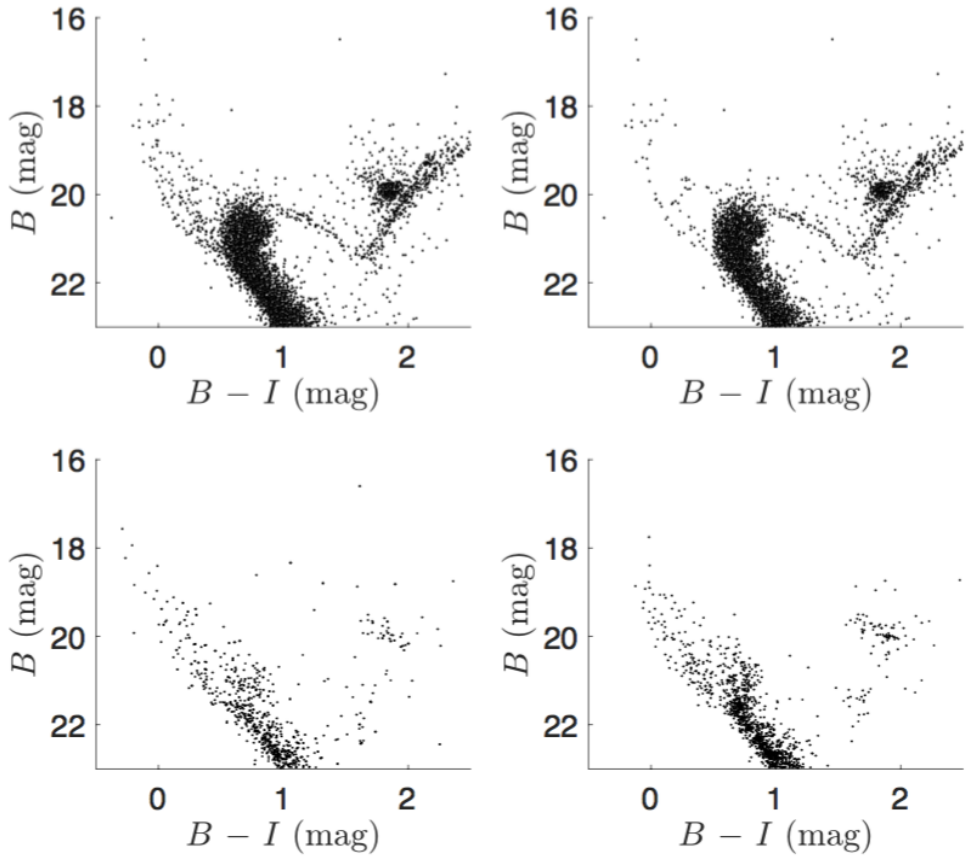

**Figure 6**: Demonstration of our field-star decontamination procedure for NGC 1783. (*top left*) Raw data; (*top right*) Decontaminated colour–magnitude diagram; (*bottom left*) Reference field-star data; (*bottom right*) Field-star-subtracted colour–magnitude diagram.

Our initial publication of these clusters with truly multiple populations triggered a study that challenged our results (Cabrera-Ziri et al. 2016), suggesting that we had merely identified the background population instead of a population of genuine cluster stars. We recently rebutted this suggestion, providing clear evidence of the reality of these younger stars as associated with their host clusters rather than the background field (Li et al. 2016c). In essence, we provided four lines of supportive evidence:

1. All younger populations in our sample clusters exhibit radial profiles that clearly peak in the cluster cores; a background field population would not be expected to exhibit any radial dependence.
2. The younger populations are tightly associated with a single young isochrone (or two single young isochrones in the case of NGC 1783); background populations would be expected to show a spread in their ages, with a characteristic *old* age.
3. We used three different 'background' field regions to statistically correct our clusters' colour–magnitude diagrams for the presence of field star contamination. In one of the background field regions, we identified the presence of a star-forming region, which hence prompted us to caution against the use of that particular section of the background field data frame. Our challengers did not exclude the region affected by active star formation, and hence their field-star subtraction inadvertently subtracted the younger cluster stars.

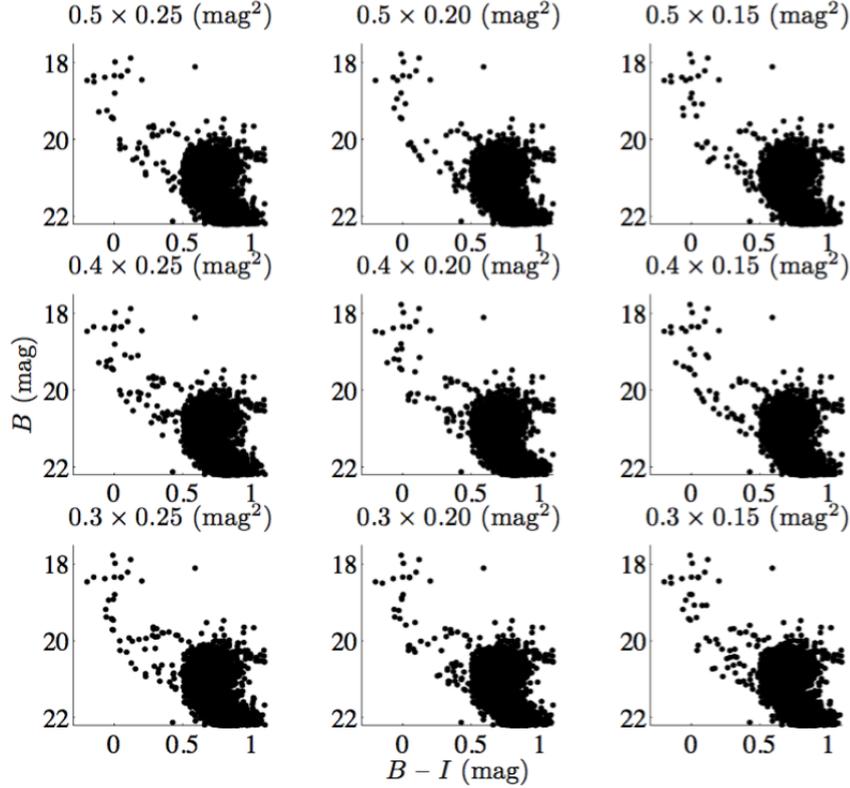

**Figure 7**: Effects of the adopted colour–magnitude cell sizes adopted (indicated for each panel separately) on our ability to retrieve the younger stellar sequences in NGC 1783.

4. We were accused of oversubtracting field stars from colour–magnitude cells that contained fewer stars than the numbers of field stars expected on statistical grounds. However, in the *Supplementary Materials* of Li et al. (2016a), we had carefully assessed the effects of varying our cell sizes, showing that for a very large range of sizes the young stellar populations remained clearly visible.

Extended Data Fig. 6 of Li et al. (2016a) illustrates the field-star decontamination process in detail. Here, in Fig. 6, we first show how our statistical field-star decontamination procedure is applied in practice, using NGC 1783 as our example. We additionally show in Fig. 7 that almost any choice of background cell size retains the morphological features of the younger populations in NGC 1783. In other words, the reality of the younger sequences in this cluster is robustly verified; only for very small or very large colour–magnitude cell sizes, respectively smaller than $0.10 \times 0.05$ mag$^2$ and larger than $1.0 \times 0.5$ mag$^2$, do the features wash out. Thus, we stand by the Li et al. (2016a) results, with the proviso that our interpretation has now been updated to reflect that cluster mergers are more likely to have led to the observed configurations than star formation resulting from gas accretion.

In summary, it has become clear that the presence of extended main-sequence turn-off regions in intermediate-age (~1–3 Gyr-old) star clusters does not necessarily imply that the clusters' member stars are characterized by a range in stellar ages. A simple stellar population containing single-aged stars but with a range in stellar rotation rates seems, at present, one of the more robust and convincing explanations. While our results do not challenge the multiple populations routinely observed in old globular clusters, at present we argue that our understanding of star cluster formation and evolution requires a significant reassessment; long-held ideas may have to be abandoned, but such is the nature of science!

**Acknowledgements**
RdG thanks the organisers of the Frontiers in Theoretical and Applied Physics | UAE 2017 meeting for their efforts to put together an interesting and high-quality conference. He would like to single out Randa Asa'd in particular for her exemplary assistance and highly competent organisational skills. RdG was partially supported by the National Natural Science Foundation of China (NSFC; grants U1631102, 11373010, and 11633005). CL acknowledges funding support from the Macquarie Research Fellowship Scheme.